\documentclass[aps,prd,reprint,twocolumn,groupedaddress]{revtex4}
\usepackage{amsmath}
\usepackage{graphicx}
\usepackage{euscript}
\usepackage{braket}

\begin{document}
\title{Hadron production in d-Au, p-Pb, Pb-Pb and Xe-Xe collisions at RHIC and LHC: onset of color glass condensate in light of  an analytical solution of BK equation }


\author{Pragyan Phukan}
\email[]{pragyanp@tezu.ernet.in}
\affiliation{HEP laboratory, Department of Physics, Tezpur University, India}
\author{Madhurjya Lalung}
\email[]{mlalung@tezu.ernet.in}
\affiliation{HEP laboratory, Department of Physics, Tezpur University, India}

\author{Jayanta Kumar Sarma}
 \email[]{jks@tezu.ernet.in}
\affiliation{HEP laboratory, Department of Physics, Tezpur University, India}


\begin{abstract}
An exact analytical solution to the Balitsky-Kovchegov equation is proposed incorporating Mellin transformation with saddle point approximation. The theoretical significance of this solution is its simplicity in form which requires very less number of parameters. Using this solution along with the idea of color glass condensate, a very good description of RHIC and LHC data on differential yield is obtained for d-Au, p-p and p-Pb collisions. Quantitative prediction for the  nuclear modification factors for p-Pb, Pb-Pb and Xe-Xe collisions relevant for very recent LHC run 2 is yielded.
\end{abstract}


\maketitle


Heavy ion collisions performed previously at RHIC \cite{9,10} and recently at LHC \cite{15,14,16} probe the color glass condensate (CGC) \cite{1} regime of QCD governed by the nonlinear effects in the colliding nuclei wave functions as well as gluon saturation. Pb-Pb collision at LHC opens up an excellent opportunity to study the hot-dense hypothetical QCD medium arguably called quark gluon plasma (QGP). Various theoretical frameworks \cite{2,3,4} that are modeled to study these QCD processes, primarily based on the concept of saturation scale ($Q_{sA}$) that characterizes the onset of nonlinear coherent phenomena. The growth of the saturation scale with increasing energy is yielded by the Balitsky-Kovchegov (BK) nonlinear evolution equation \cite{19,20}. This portrays a direct link between the gluon saturation dynamics and the experimentally measured hadron yields.

\par The BK equation in position space is characterized by the dipole scattering amplitude, $\EuScript{N}(r,Y)$ on a target, where the target is specified by the initial condition. For large and homogeneous  targets, Fourier transformation of the scattering amplitude to the momentum space yields
\begin{equation}
\label{1}
\partial_{Y}\EuScript{N}=\bar{\alpha}_s\chi(k^2) \EuScript{N}-\bar{\alpha}_s \EuScript{N}^2,
\end{equation}
where $\bar{\alpha}_s=N_c\alpha_s/\pi$ and the quantity $\EuScript{N}(k,Y)=\int_{0}^{\infty}\frac{d^2r}{2\pi r^2}e^{i r.k}\EuScript{N}(r,Y)$ is related to unintegrated gluon distribution (UGD) of the target. The operator  $\chi$ is the well known LL BFKL kernel in momentum space.  The nonlinear term in \eqref{1} is not acted on by the kernel operator and thereby posses a particularly simple representation. 
\par To solve the BK equation \eqref{1} in Mellin space we recast \eqref{1} in terms of its inverse function $\underline{\EuScript{N}}=\EuScript{N}^{-1}$ such that the semilinear PDE transforms into a linear one and then diagonalize the same by Mellin transformation
\begin{equation}
\label{2}
\partial_{Y}\tilde{\underline{\EuScript{N}}}=-\bar{\alpha}_s\tilde{\chi}(\gamma) \tilde{\underline{\EuScript{N}}}-2\pi \bar{\alpha}_s\delta(i\gamma),
\end{equation}
where $\delta(i\gamma)$ is a Dirac-delta function. The Mellin transformation of the scale invariant kernel $\chi$ in \eqref{2} can be formulated in terms of the first order harmonic number $\EuScript{H}_{-\gamma}^{(1)}$ as follows,
\begin{equation}
\begin{aligned}
\label{3}
\tilde{\chi}(\gamma)&=\int_{0}^{\infty}\frac{du}{u}\left[\frac{u^{-\gamma+1}-1}{|u-1|}+\frac{1}{\sqrt{4u^2+1}}\right]\\
&=\pi\cot(\pi \gamma)-2\EuScript{H}_{-\gamma}^{(1)}
\end{aligned}
\end{equation}
implies,
\begin{equation}
	\label{3a}
\tilde{\chi}'(\gamma)=-\pi^2\csc(\pi\gamma)^2+\pi^2/3-2\EuScript{H}_{-\gamma}^{(2)}.	
\end{equation}
Inverse Mellin transformation on the solution of \eqref{2} yields
\begin{equation}
\label{4}
\begin{aligned}
\underline{\EuScript{N}}&=\frac{1}{2\pi i}\int_{c-i\infty}^{c+i\infty} d\gamma (k^2/k_0^2)^\gamma \times\\
&\left[\frac{2\pi\bar{\alpha}_s\delta(i\gamma)}{\tilde{\chi}(\gamma)}+e^{\tilde{\chi}(\gamma)(y_0-Y)}\left(\tilde{\underline{\EuScript{N}}}_0-\frac{2\pi\bar{\alpha}_s\delta(i\gamma)}{\tilde{\chi}(\gamma)}\right)\right],
\end{aligned}
\end{equation}
with $\tilde{\underline{\EuScript{N}}}_0=\tilde{\underline{\EuScript{N}}}(k,y_0)$ where the initial condition $y_0$ is chosen large enough to ensure the validity of BK equation. An arbitrary scale $k_0$ is introduced for dimensional purpose. Note that $\tilde{\chi}(\gamma)$ is symmetric around $\gamma=1/2$ and along the contour $\gamma=1/2+i\nu$, with $-\infty<\nu<\infty$ has its maximum at the "saddle" point  $\gamma=c=1/2$. This indicates that the $\gamma\sim1/2$ region dominates the small-$x$ behavior of the $N(k,Y)$ which seem to recover the familiar LL BFKL characteristic function behavior. To produce the iterative solution of \eqref{2} we therefore set $\gamma=1/2+i\nu$ in \eqref{4} and expand $\tilde{\chi}(\gamma)$ about $\nu=0$ i.e.  $\tilde{\chi}(1/2+i\nu)=\lambda -1/2\Omega\nu^2+\EuScript{O}(\nu^2)$ with $\lambda=\bar{\alpha}_s4\ln2$ and $\Omega=\bar{\alpha}_s28\zeta(3)$, $\zeta$ being Reimann zeta function. We also need to expand $\tilde{\underline{\EuScript{N}}}(1/2+i\nu,y_0)$ around $\nu=0$,

\begin{equation}
\label{5}
\tilde{\underline{\EuScript{N}}}(1/2+i\nu,y_0)\approx \tilde{\underline{\EuScript{N}}}(1/2,y_0)e^{\frac{\nu\partial \ln\tilde{\underline{\EuScript{N}}}(1/2+i\nu,y_0)}{\partial \nu}|_{\nu=0}}.
\end{equation}
Substituting these expansions back to \eqref{4} and performing $\nu$ integration along with various Dirac-delta operations  one may arrive at
\begin{equation}
	\label{6}
	\begin{aligned}
	\underline{\EuScript{N}}=& \frac{(k^2/k_0^2)^\frac{1}{2}e^{\lambda(y_0-Y)}\tilde{\underline{\EuScript{N}}}(1/2,y_0)}{(2\pi)^{\frac{1}{2}}\sqrt{\Omega| y_o-Y|}}e^{-\frac{\ln^2(k^2/\tilde{k}_s^2)}{2\Omega\ln(y_0-Y)}}\\
	&+\frac{\bar{\alpha}_s}{\lambda+1/8\Omega}\{1-e^{(\lambda+1/8\Omega)(y_0-Y)}\},
	\end{aligned}	
\end{equation}
where we define an arbitrary scale $\ln\tilde{k_s}^2=i\frac{d}{d\nu}[\ln \underline{\tilde{\EuScript{N}}}(y_0,1/2+i\nu)]|_{\nu=0}$. Above equation serves as the analytical solution of \eqref{2}, however, the original solution $\EuScript{N}$ of the BK equation \eqref{1} would be the immediate reciprocal of \eqref{6} as we recall $	\underline{\EuScript{N}}=\EuScript{N}^{-1}$ from our assumption.
\par A few comments on the solution \eqref{6} are in order. The first term of \eqref{6} forecasts the characteristic $\sim e^{\lambda Y}$ behavior of $\EuScript{N}$, so does UGD, similar to LL BFKL evolution. On the other hand, the second term modulates the power growth arising from the linear contribution thereby suppressing the gluon evolution. Interestingly, at pre asymptotic rapidities ($Y\sim y_0$), the nonlinear contribution vanishes since the driving term
\begin{align*}
	e^{(\lambda+1/8\Omega)(y_0-Y)}|_{Y\rightarrow y_0}\rightarrow 1.
\end{align*}
Thus at not so large $Y$, the BK solution \eqref{6} recovers usual linear BFKL behavior. Surprisingly the nonlinear contribution in the solution is irrelevant of the momentum scale, $k^2$ and it depends particularly on the rapidity only.

\par Equation \eqref{6} suggests the exact asymptotic solution of the BK equation and thus geometrical scaling appears as a universal property of these kinds of nonlinear equations. However, the detailed phenomenological studies on the gluon evolution in context of geometrical scaling is beyond the scope of this letter. Instead, in this work, we try to address a more realistic aspect of the solution relating our theory to the ongoing collider physics experiments.


\par In high energy asymmetric collisions, there are at least two particular approaches leading to hadron production. The first one is the $k_t$ factorization formalism, appropriate for central rapidity region, where both the target and the projectile are characterized in terms of UGD. But at more forward rapidities, $k_t$ factorization fails to grasp the dominant contribution to the cross section coming from valence quarks of the projectile. In this situation, so called \textit{hybrid} formalism is most pronounced where the large-$x$ degrees of freedom of the projectile is characterised by usual collinear gluon distribution. In hybrid formalism, the differential cross section for forward hadron production with transverse momentum, $p_t$ and pseudo rapidity $\eta$, in proton-nucleus collisions \cite{5,6}  is given by 

\begin{equation}
\label{7}
\begin{aligned}
\frac{d N_h^{pA\rightarrow hX}}{d \eta d^2 p_t}=&\frac{C}{(2\pi)^2}\sum_{q}\int_{x_F}^{1}\frac{d z}{z^2}\bigg[\frac{x_F}{z}f_{q/p}(x_F/z,p_t^2)\\
&\times \tilde{N}_F(x',p_t^2)D_{h/q}(z,p_t^2)\\
&+\frac{x_F}{z}f_{g/p}(x_F/z,p_t^2)\tilde{N}_A(x',p_t/z)D_{h/g}(z,p_t^2)
\bigg]
\end{aligned}
\end{equation}
where the kinematics are  $x_F=\sqrt{m_h^2+p_t^2}/\sqrt{S_{NN}}$ and $x'=x_Fe^{-2y_h}/z$ with $\sqrt{S_{NN}}$ being the collision energy per nucleon. For highly relativistic particles the difference between rapidity $y_h$ and pseudorapidity $\eta$ can be neglected, $\eta\approx y_h$. The projectile is characterized by the collinear PDFs $f_{i/p}$, however its partons eventually acquire a large transverse momentum $p_t$ after interaction with the small-$x$ fields of the saturated nucleus causing multiple scattering which is described by the UGDs in the adjoint or fundamental representation $\tilde{N}_A(F)$. The term $D_{h/i}$ refer to the final-state hadron fragmentation function which characterizes the hadronization of the scattered parton. The lack of impact parameter dependence in the calculation is accounted for by the normalization factor C.
\begin{figure}
	\includegraphics[width=0.8\linewidth]{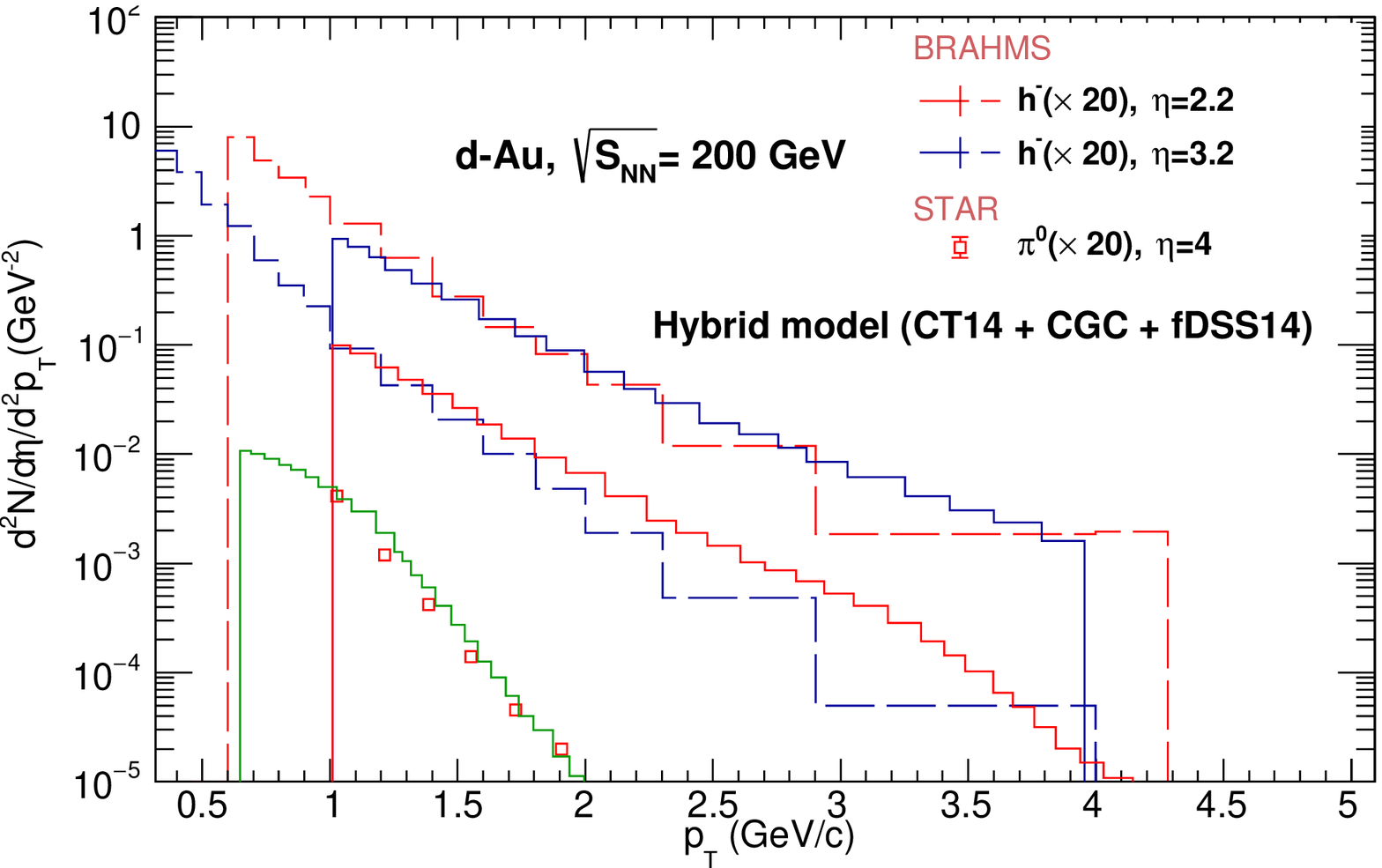}
	\includegraphics[width=0.8\linewidth]{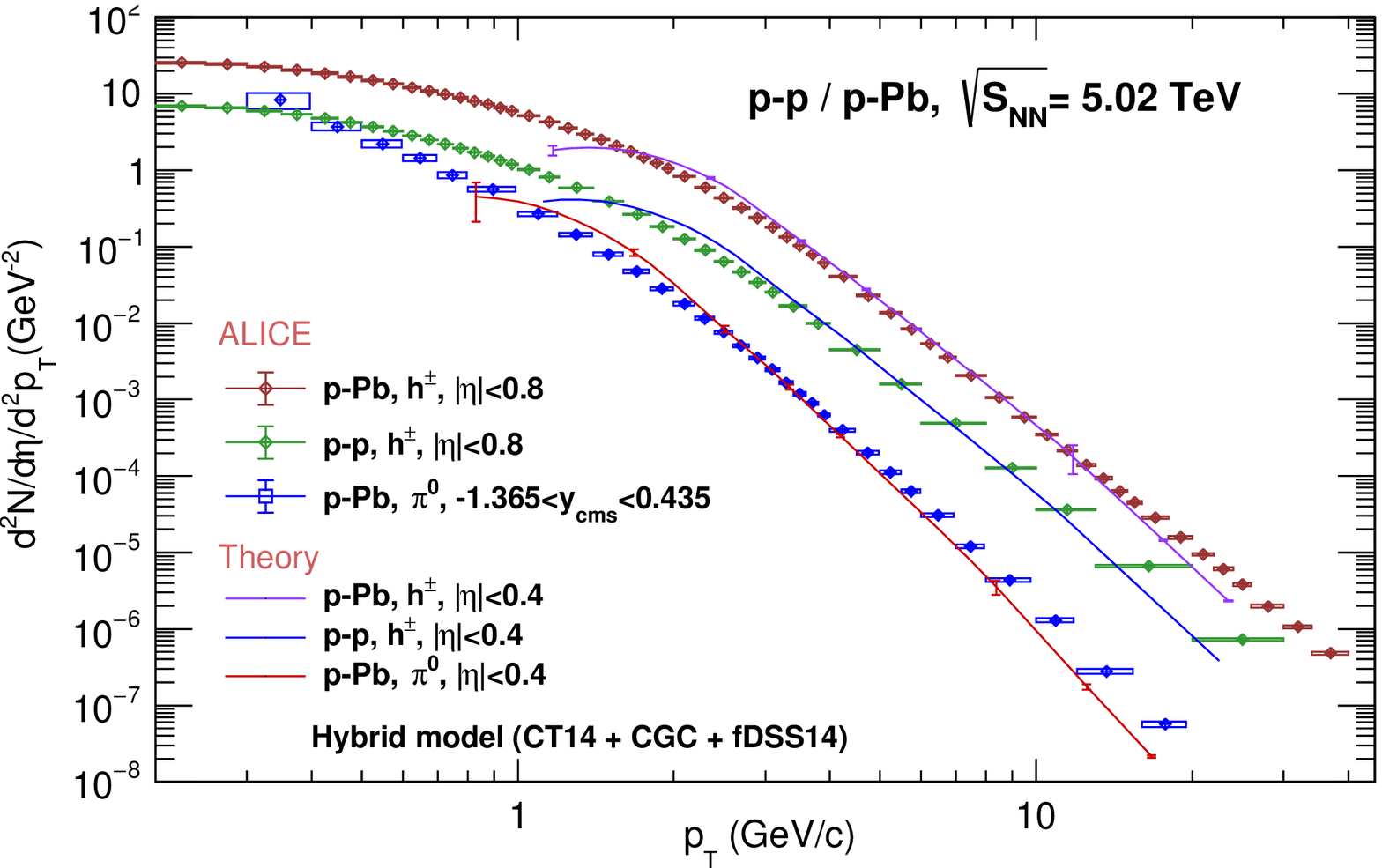}
	\caption{ Charged hadrons $h^{-}$ / $h^{\pm}$ and $\pi^0$ yields in (a) deuteron-gold collision at RHIC ($\sqrt{S_{NN}}=200~\text{GeV}$) and (b) proton-proton, proton-lead collison at LHC ($\sqrt{S_{NN}}=5.02~\text{TeV}$). Solid lines are theoretical predictions. Data by BRAHMS, STAR and ALICE collaboration.}
	\label{f1}
\end{figure}
\par The nuclear UGDs,  $\tilde{N}_{F(A)}$ entering \eqref{7} are given by the two dimensional Fourier transform of the imaginary part of the forward dipole scattering amplitude in the fundamental (F) and adjoint (A) representation
\begin{equation}
\label{8}
\begin{aligned}
\tilde{N}_{F(A)}(x,k)=\int d^2 \text{r}e^{-i\text{ k.r}}(1-\EuScript{N}_{F(A)}(r, Y)),
\end{aligned}
\end{equation}
where $r$ is the dipole size and rapidity $Y=\ln(x_0/x)$. The small-$x$ dynamics of the dipole amplitudes $\EuScript{N}_{F(A)}(r, Y)$ in \eqref{8} is given by the solution of BK equation \eqref{6} (inverse) but in momentum space, however, it is straightforward to switch between position and momentum domains via Fourier transform. The initial conditions for the evolution of dipole-nucleus amplitude are taken from the semiclassical Mclerran-Venugopalan (MV) model \cite{17},
\begin{equation}
\label{9}
\tilde{\EuScript{N}}_A(r,Y=0)=1-\text{exp}\bigg[-\frac{r^2 Q_{s0}}{4}\ln\left(\frac{1}{\Lambda r}+e\right)\bigg]
\end{equation}
where $Q_{s0}$ is the initial saturation scale probed by quarks. The constant $e$ serves as an infrared regulator and $\Lambda\approx0.24$~GeV. The scattering amplitude in fundamental representation, $\EuScript{N}_F$ can also be parametrized as in \eqref{9} but with the replacement $Q_s^2\rightarrow Q_s^2 C_F/C_A=4/9 Q_s^2$ \cite{18}. 
\begin{figure}[h]
	\includegraphics[width=0.8\linewidth]{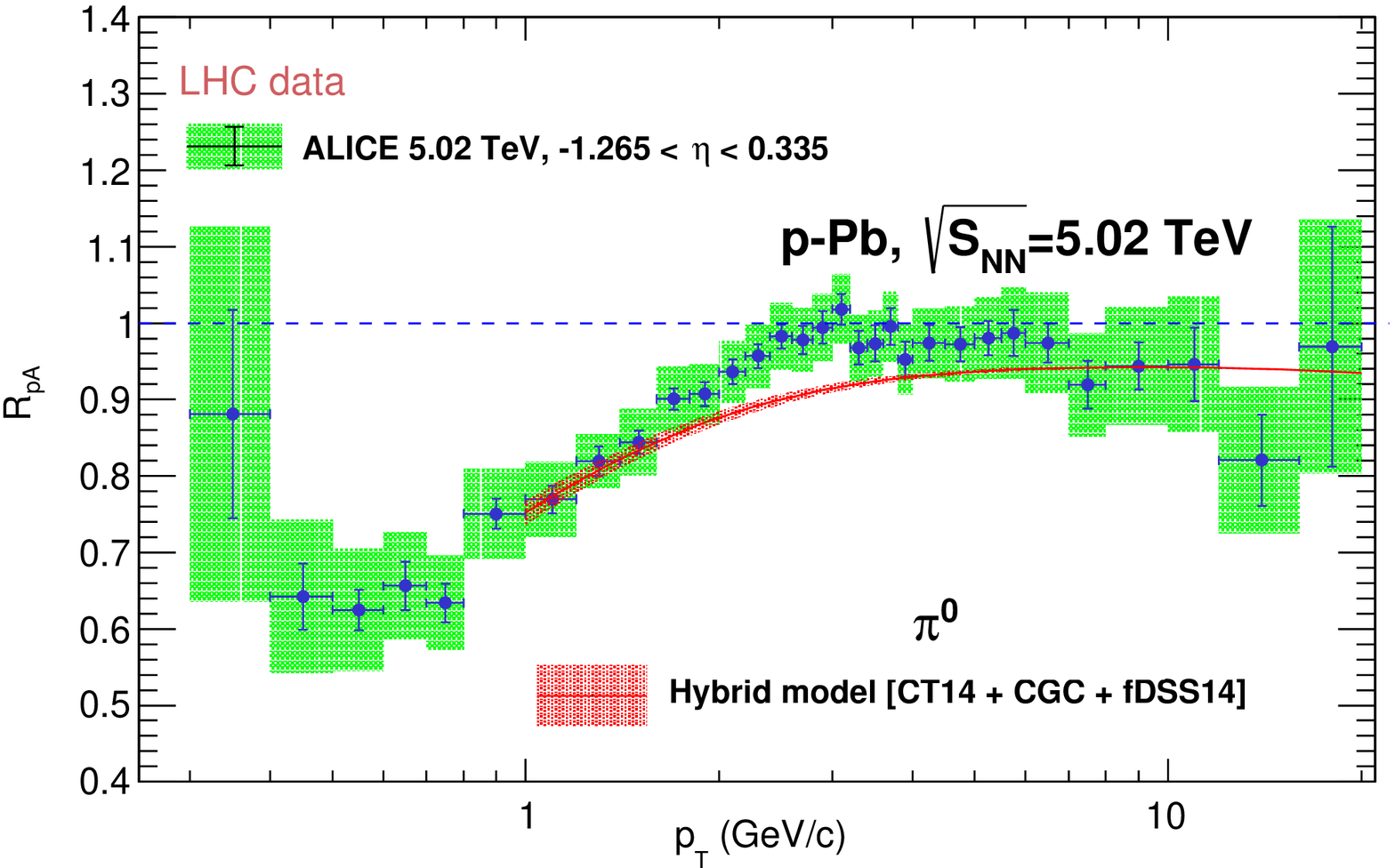}
	\includegraphics[width=0.8\linewidth]{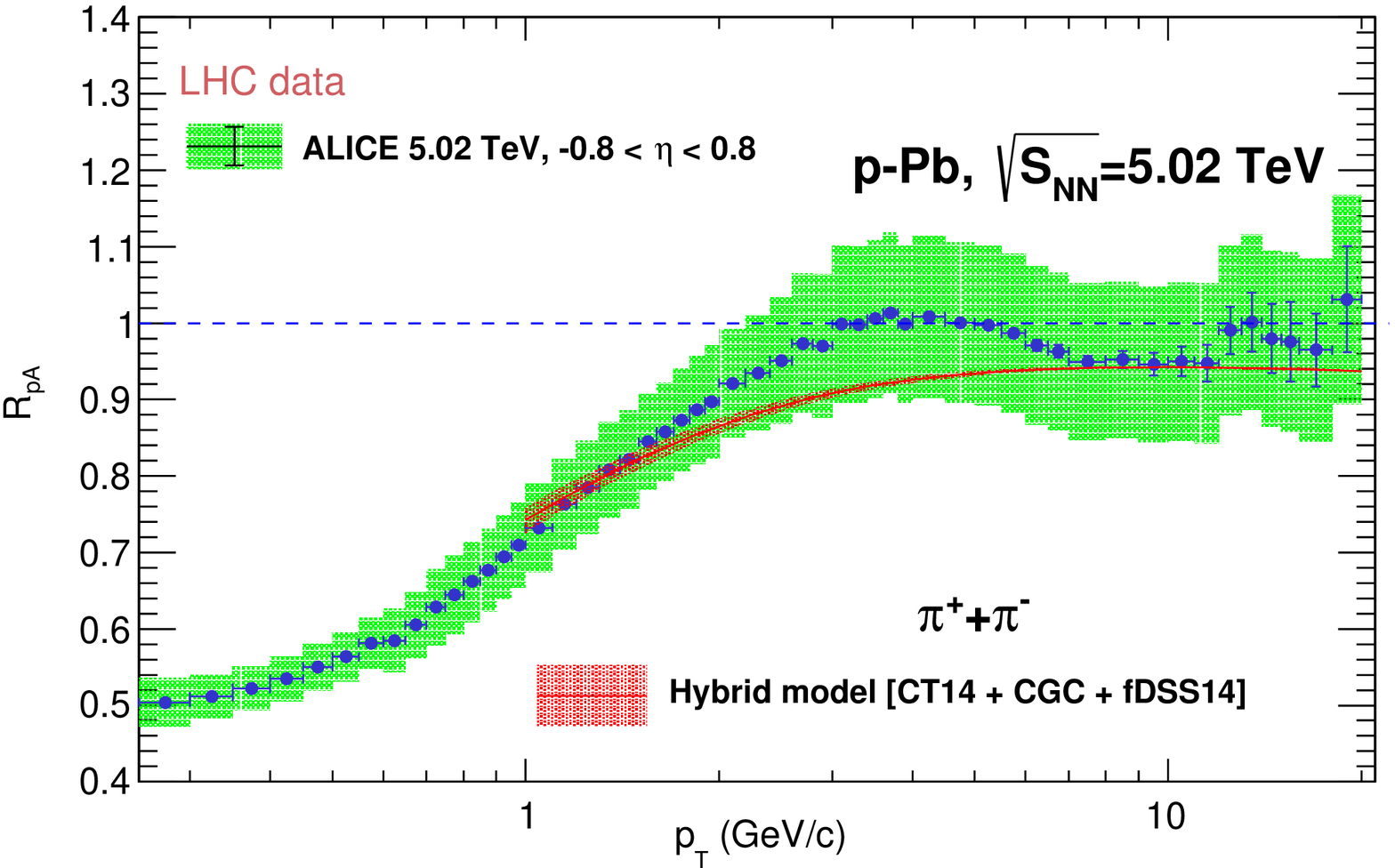}
	\caption{ Neutral pion ($\pi^0$) and charged pions ($\pi^+,\pi^-$) nuclear modification factors for Pb-Pb collision at $\sqrt{S_{NN}}=5.02~\text{TeV}$.}
	\label{f2}
\end{figure}
\par With this setup, we obtain a very good description of RHIC and LHC run 2 data for hadron yields of different hadron species in p-p, d-Au and p-Pb collisions. The invariant differential yields in d-Au collision at RHIC energy $\sqrt{S_{NN}}=200$~GeV and pseudo-rapidity $\eta$ = 2.2 and 3.2 for negatively charged hadrons ($h^-$) (BRAHMS collaboration \cite{9}) and $\eta$ = 4 for neutral pions ($\pi^0$) (STAR collaboration \cite{10}) is shown in Fig.~\ref{f1}. In analogy to \cite{11}, we obtain a quark saturation scale $Q_{s0}\approx0.43~\text{GeV}^2$ for gold nucleus at $x_0=0.01$. Corresponding gluon saturation scale $Q_{s0}^2$ 	is found to be $Q_{s0}^2\approx0.97~\text{GeV}^2$. For entire calculations performed with hybrid formalism, we adopt the DSS NLO fragmentation functions \cite{8} and CT14 NLO collinear PDFs \cite{7} as default. Our results show a good agreement with RHIC data. However, for the better description of neutral pion data, a normalisation factor $C=0.6$ is multiplied, while charged hadron yields are sketched without any normalisation (i.e. C=1). Considering the sizable uncertainty in the data, the precise value of the normalisation constant does not have much meaning. In Fig.~\ref{f1}, the $\pi^0$ and $h^{\pm}$ yields for p-Pb collisions and $h^{\pm}$ yield for p-p collisions measured in ALICE at $\sqrt{S_{NN}}=5.02~\text{TeV}$ \cite{14} are also shown. Corresponding pseudo rapidity windows for $h^{\pm}$ and $\pi^0$ are $|\eta|<0.8$ and $|\eta|<0.4$ respectively. Here for all calculation the normalisation is set to 1. The theoretical  measurement are clearly consistent within the uncertainties over the entire $p_t$ range.
\par The modification of forward particle production is quantified with the nuclear modification factor,
\begin{equation}
\label{10}
R_{pA}=\frac{d^2 N^{pA}/d\eta d p_t}{\braket{T_{pA}} d^2 \sigma ^{pp}/d\eta d p_t},
\end{equation}
where $N^{pA}$ is the charged particle yield in pA collisions and $\sigma^{pp}$ is the particle production cross-section in p-p collision. The average nuclear overlap function, $\braket{T_{pA}}=\braket{N_{coll}}/\sigma_{inel}^{NN}$ depends on the collision centrality determined from the Glauber model \cite{12}. For instance, at $\sqrt{S_{NN}}=5.02$~TeV, $\braket{T_{pPb}}\approx0.098 \quad \text{mb}^{-1}$ for minimum bias. Note that computing the ratio \eqref{10} removes sensitivity to the normalization factor C. Our $R_{pPb}$ calculations scaled by $\braket{T_{pPb}}$ are displayed in Fig.~\ref{f2}  at $\sqrt{S_{NN}}=5.02~\text{TeV}$ subject to the pseudo rapidity windows $-0.8<\eta<0.8$ for neutral pions ($\pi^0$) and $-1.265<\eta<0.335$  for charged pions ($\pi^++\pi^-$) agree well with the ALICE data within the uncertainties over the complete $p_t$ domain. Both theory and data indicate a sizable suppression around 30\% at $p_t=1~\text{GeV}$, whereas the ratio $R_{pPb}$ is seen tending to  unity for $p_t$ above 2 GeV for both neutral pions and charged pions.

\begin{figure}[h]
	\includegraphics[width=0.48\linewidth]{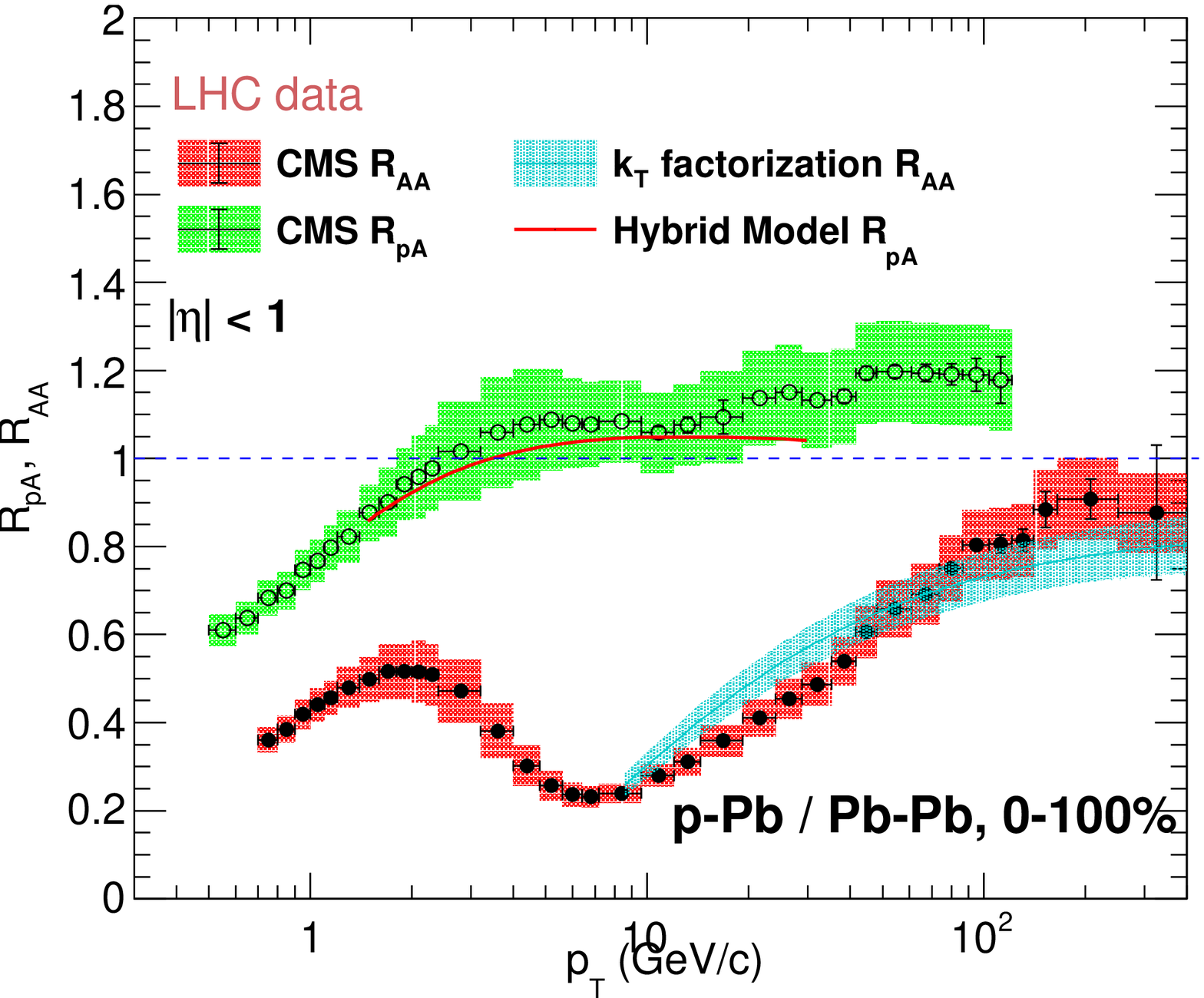}
	\includegraphics[width=0.48\linewidth]{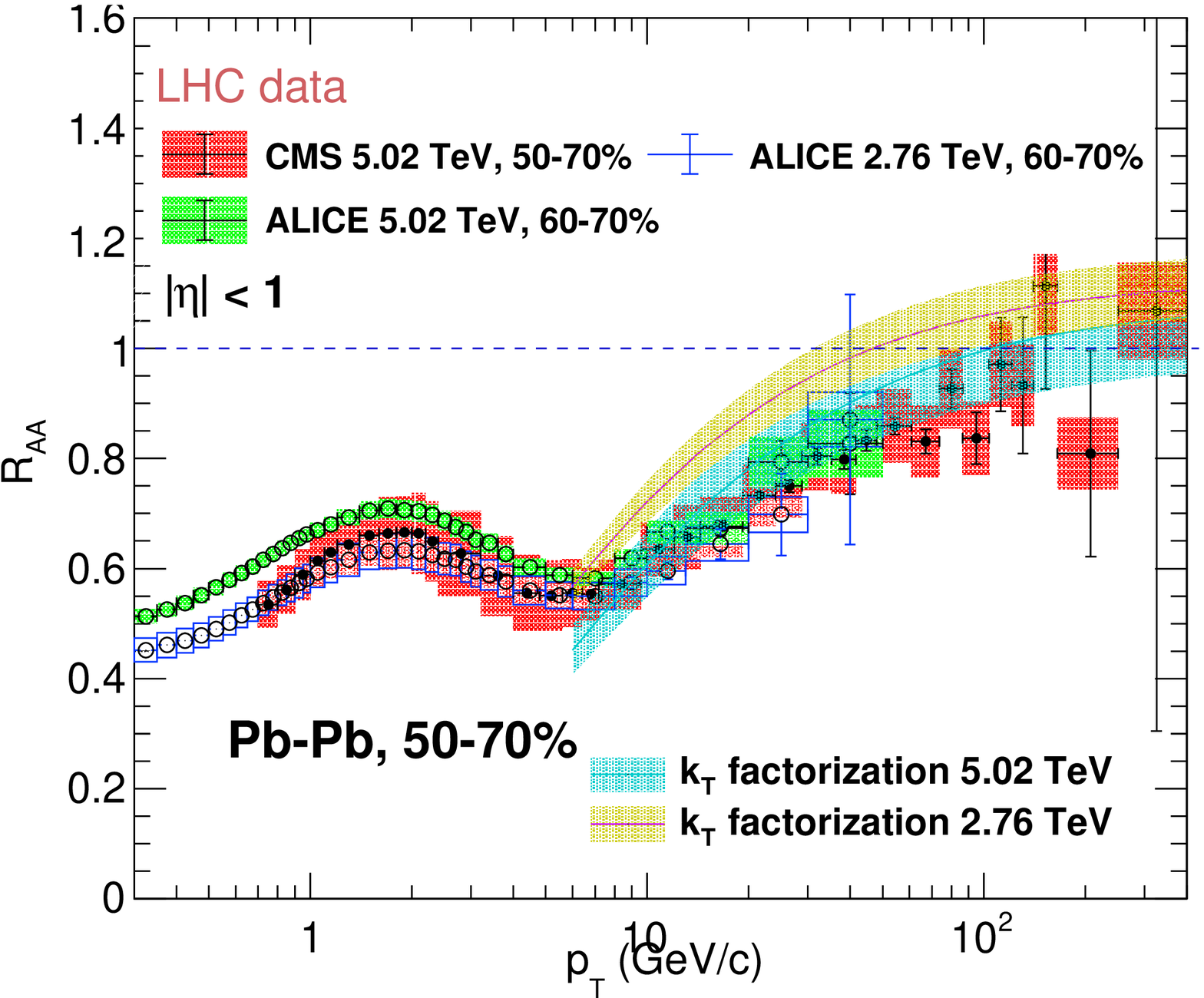}
	\includegraphics[width=0.48\linewidth]{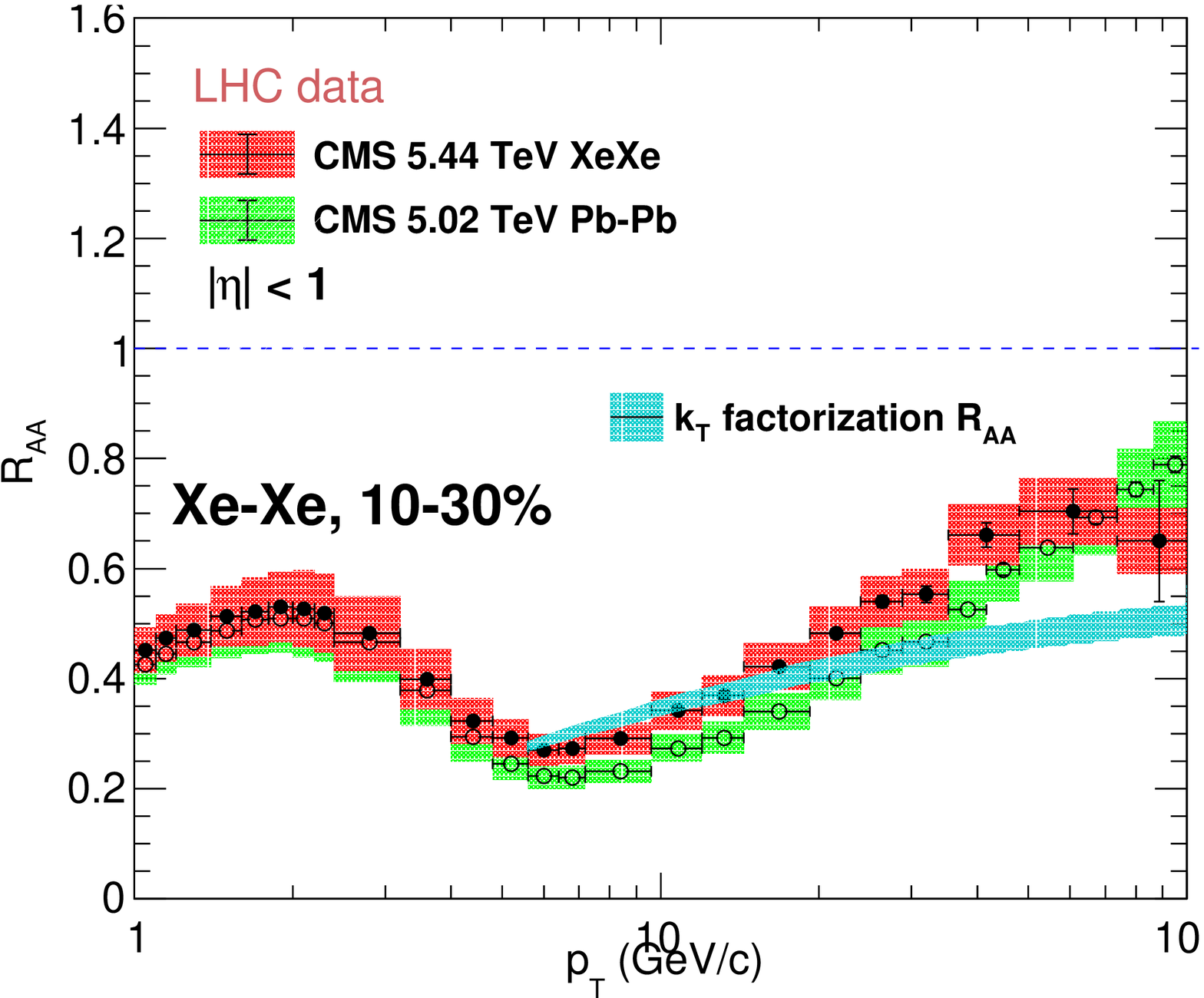}
	\includegraphics[width=0.48\linewidth]{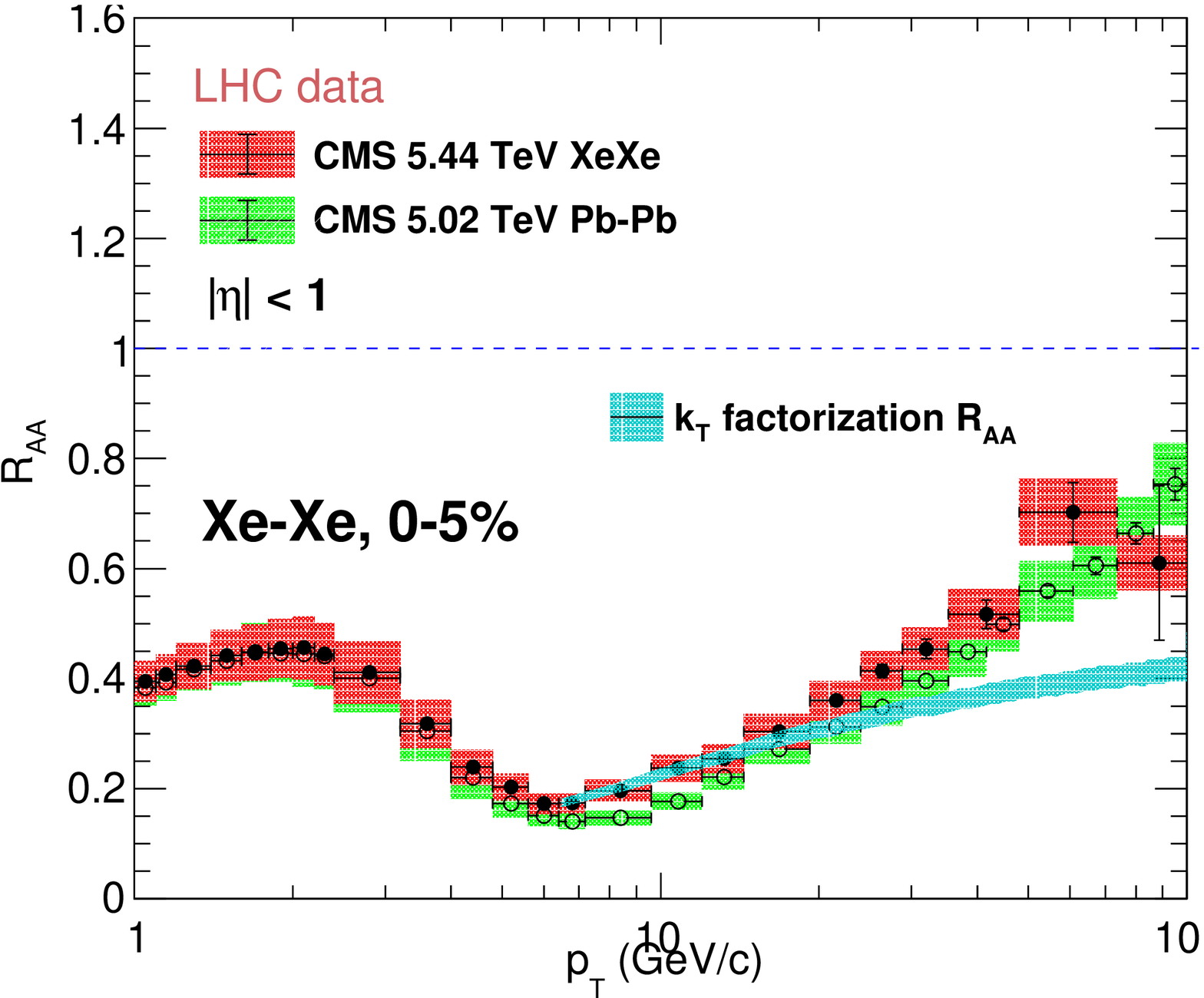}	
	\caption{ Charged hadron $R_{AA}$ for (a) Pb-Pb collisions at $\sqrt{S_{NN}}=5.02~\text{TeV}$ for centrality bins 0-100\%, 50-70\% and (b) Xe-Xe collision at $\sqrt{S_{NN}}=5.44~\text{TeV}$ for centrality bins 10-30\%, 0-5\%. }
	\label{f3}
\end{figure}

\par We have also checked our BK analytical solution + CGC framework to testify charged hadron production particularly at mid rapidities using $k_t$ factorization formalism. At mid rapidities, $x_F/z$ becomes too small to valid hybrid formalism in \eqref{7}. The $k_t$ factorization formula is valid for small values of $x$, for both the projectile and targets. But since it includes only gluonic degrees of freedom, this formalism is insufficient at very forward rapidities or large $p_t$. Within $k_t$ factorization approach \cite{13} the pseudo rapidity density of charged particle can be calculated via.
\begin{equation}
\label{11}
\frac{d N^{ch}}{dy dp_t}=\frac{2\pi  K}{C_F}\frac{\alpha_s}{p_t^2}\int dk_T^2\phi_{A_1}(x_1,k_t^2)\phi_{A_2}(x_2,(p-k)_t^2),
\end{equation}
where the variables $x_{1,2}=p_t/\sqrt{S_{NN}}e^{\pm y}$ and $K$ is accounted for the normalisation purpose. The UGD $\phi$, in \eqref{11} is related to collinear gluon distribution, $\phi(x,k)\propto \frac{d (xG(x,k^2))}{d^2k}$ and is given in terms of the dipole scattering amplitude
\begin{equation}
\label{12}
\phi(y,k)=\int\frac{d^2r}{2\pi r^2}e^{ir.k}\EuScript{N}(y,r).
\end{equation}

\par The CMS and ALICE data \cite{15,16} for the nuclear modification factor $\text{R}_{\text{AA}}$ for primary charged particles in Pb-Pb ($\sqrt{S_{NN}}=5.02~\text{TeV}$) and Xe-Xe ($\sqrt{S_{NN}}=5.44~\text{TeV}$) collisions are shown in Fig.~\ref{f3} for the centrality bins 0-100\%, 50-70\%, 10-30\% and 0-5\%. For minimum bias (0-100\%), the ratio $R_{pA}$ is also included in Fig.~\ref{f3} to sketch a quantitative comparison with $R_{AA}$. The pronounced suppression in our theoretical prediction is seen at $p_t<2~\text{GeV}$ for p-Pb system and at $8< p_t<100~\text{GeV}$ for Pb-Pb system which is consistent with CMS data. In the peripheral (50-70\%) bin, $R_{AA}$ measurements at $\sqrt{S_{NN}}=2.76~\text{TeV}$ is also shown to examine the sensitivity of the ratio towards probe energy. The measured distribution $R_{AA}$ at 2.76~TeV and 5.02~TeV are found to be quantitatively similar at intermediate $p_t$, but above about 10~GeV, $R_{AA}$ at 5.02~TeV tend to be slightly smaller than the same at 2.76~TeV. In Fig.~\ref{f3}, $R_{AA}$ for Xe-Xe collision at $\sqrt{S_{NN}}=5.44~\text{TeV}$ is also sketched for two centrality bins 0-10\% and 10-30\%. The theoretical prediction seem to be in rough agreement with CMS data at moderate $p_t$, however, towards high $p_t$, our theory tend to predict much stronger suppression compared to the data. From this analysis it is clear that our theory holds good  for minimum bias measurements, while it might be inappropriate for most central events. This is expected as our theory is impact parameter $b$ independent and thus one have to improve the theory for the $b$ dependence of the UGD to study different centrality bins which is beyond the scope of this letter.
\par In summary, we have proposed an analytical solution to BK equation incorporating Mellin transformation with saddle point approximation. The main key feature of this solution is its simplicity in form with relatively very less number of parameters, which might be handy for investigation of various characteristics of such parton evolution processes. At pre asymptotic $y$, the solution is found to reconstruct the characteristics linear BFKL solution. The feasibility of the solution is tested at both RHIC and LHC energies. As per the good consistency between the theory and data is concerned, the analytical solution of BK equation could be a very reliable framework for exploring quark gluon plasma at LHC as well as future colliders.

\bibliography{111.bib}

\end{document}